\documentclass{article}

\usepackage{PRIMEarxiv}
\usepackage[english]{babel}
\usepackage{natbib}
\usepackage{amsmath,amssymb,amsfonts}
\usepackage{algorithmic}
\usepackage{graphicx}
\usepackage{textcomp}
\usepackage{xcolor}
\usepackage{booktabs}
\usepackage{subfig}
\usepackage{hyperref}
\usepackage{caption}
\begin{document}

\tolerance=999
\sloppy

\title{
  MIA-3DCNN: COVID-19 Detection Based on a 3D CNN
}

\author{
  Igor Kenzo Ishikawa Oshiro Nakashima*,
  Giovanna Vendramini*,
  Helio Pedrini \\
  Institute of Computing, University of Campinas \\
  Campinas-SP, Brazil, 13083-852 \\
  \texttt{i217967@dac.unicamp.br, g173304@dac.unicamp.br, helio@ic.unicamp.br}
}
\footnote{* These authors contributed equally to this work.}
\maketitle

\begin{abstract}
Early and accurate diagnosis of COVID-19 is essential to control the rapid spread of the pandemic and mitigate sequelae in the population. Current diagnostic methods, such as RT-PCR, are effective but require time to provide results and can quickly overwhelm clinics, requiring individual laboratory analysis. Automatic detection methods have the potential to significantly reduce diagnostic time. To this end, learning-based methods using lung imaging have been explored. Although they require specialized hardware, automatic evaluation methods can be performed simultaneously, making diagnosis faster. Convolutional neural networks have been widely used to detect pneumonia caused by COVID-19 in lung images. This work describes an architecture based on 3D convolutional neural networks for detecting COVID-19 in computed tomography images. Despite the challenging scenario present in the dataset, the results obtained with our architecture demonstrated to be quite promising.
\end{abstract}

\keywords{COVID-19 \and Deep Learning \and Convolutional Neural Networks \and Lung Images}

\section{Introduction}

The first transmissions of a new coronavirus, SARS-CoV-2 (Severe Acute Respiratory Syndrome Coronavirus 2)~\citep{cascella2022features}, occurred at the end of 2019, being identified in the region of Wuhan, China, causing the pandemic of COVID-19 during the following years. The symptoms of COVID-19 can range from none to severe. Among the aggravations of the disease is the severe pneumonia that the infection can cause, potentially allowing its detection through lung images of the infected individual.

The 3rd COVID-19 Competition~\citep{white_paper, kollias2022ai, Kollias_2021_ICCV}) is an annual challenge that encourages research in the analysis of medical lung images for the detection of COVID-19. This competition uses the COV19-CT-DB database~\citep{arsenos2022large}, containing CT scans of patients with and without COVID-19, collected between September of 2020 and November of 2021.

Each computed tomography present in this database is a three-dimensional image, represented by slices, and the number of slices per tomography varies between 50 and 700, according to specifications given at the time of performing the image exam.

The annotation of each slice was performed by four professionals, radiologists and pulmonologists, with great experience in the area, with 98\% agreement between the specialists during the annotation of the classes. The dataset was then separated into training, validation and test sets, with only the first two available to participants to be used during network training, and the last one for participants to perform inference and evaluate their methods.

The competition consists of two challenges:

\begin{enumerate}

\item \textbf{COVID Detection:} Challenge that aims to classify lungs between COVID and non-COVID classes. The dataset for this challenge is unbalanced, having 922 CT scans affected by COVID-19, and 2110 healthy ones, while in the validation set, these values are 225 and 489, respectively. Figure~\ref{fig:miadiagram} shows a diagram of the layout of the images made available for this challenge, whereas Figure~\ref{fig:mia-class-covid} illustrates some examples of images of the COVID and non-COVID classes.

\begin{figure}[!htb]
\centering
\includegraphics[width=0.92\textwidth]{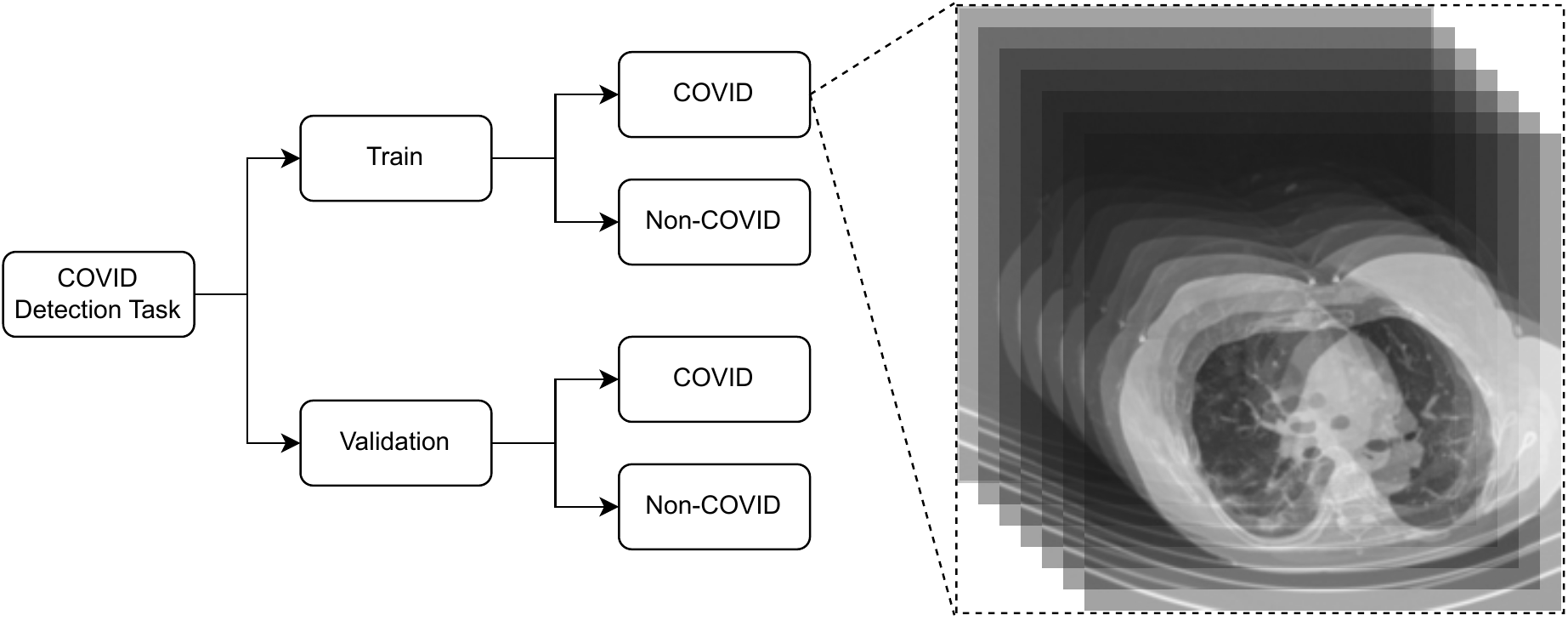}
\caption{Classification task dataset directories.}
\label{fig:miadiagram}
\end{figure}
  
\begin{figure}[!htb]
\centering
\subfloat[]{\includegraphics[width=3.5cm]{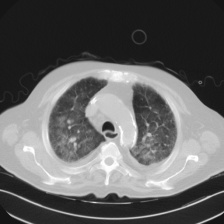} \label{fig:mia-0020-cov}} \hspace*{0.2cm}
\subfloat[]{\includegraphics[width=3.5cm]{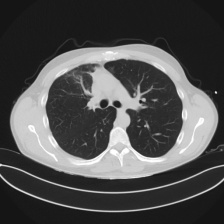} \label{fig:mia-0026-cov}} \hspace*{0.2cm}
\subfloat[]{\includegraphics[width=3.5cm]{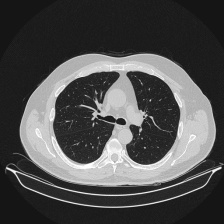} \label{fig:mia-0022-ncov}} \hspace*{0.2cm}
\subfloat[]{\includegraphics[width=3.5cm]{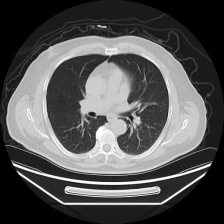} \label{fig:mia-0027-ncov}}
\caption{Examples of CT images from the MIA-COV19D database. Each of the images corresponds to one of the tomography slices, of lungs affected (for instance, (a) and (b)) or not by COVID-19 (for instance, (c) and (d)).}
\label{fig:mia-class-covid}
\end{figure} 

\item \textbf{Severity Classification:} Challenge to classify the involvement of lungs affected by COVID-19. This has the classes \textit{mild}, \textit{moderate}, \textit{severe} and \textit{critical}. The sampling for each of these groups can be seen in Table~\ref{table:mia-sev-imgs/class}, and an example image for each of the classes is presented in Figure~\ref{fig:mia-class-sev}.

\begin{table}[!htb]
\centering
\caption{Sampling for the challenge of classifying the severity of lung involvement with COVID-19. Number of images for each class (mild, moderate, severe and critical), for the training and validation sets.}
\begin{tabular}{ccc}
\toprule
\textbf{Severity} & \textbf{Training} & \textbf{Validation} \\
\midrule
Mild & 133 & 31 \\ 
Moderate & 124 & 20 \\ 
Severe & 166 & 45 \\ 
Critical & 39 & 5 \\
\bottomrule
\end{tabular}
\label{table:mia-sev-imgs/class}
\end{table}
    
\begin{figure}[!htb]
\centering
\subfloat[Mild]{\includegraphics[width=3.5cm]{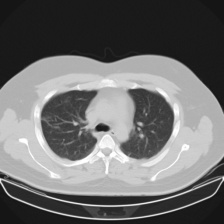} \label{fig:mia-mild}} \hspace*{0.2cm}
\subfloat[Moderate]{\includegraphics[width=3.5cm]{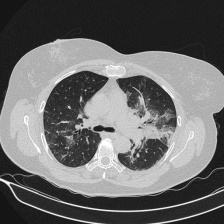} \label{fig:mia-moderate}} \hspace*{0.2cm}
\subfloat[Severe]{\includegraphics[width=3.5cm]{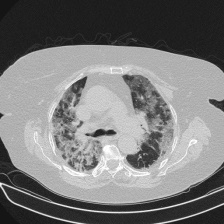} \label{fig:mia-severe}} \hspace*{0.2cm}
\subfloat[Critical]{\includegraphics[width=3.5cm]{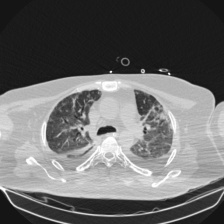} \label{fig:mia-critical}}
\caption{Each of the images corresponds to one of the tomography slices, with varying degrees of lung involvement by COVID-19.}
\label{fig:mia-class-sev}
\end{figure}

\end{enumerate}

\section{Related Work}

Recent approaches of the literature have been developed in this field and obtained great results. Some remarkable methods, which had great results in past, are the Latent Representation Analysis~\citep{kollias2020deep, kollias2020transparent}, the Spatiotemporal Feature Learning Based on Two-Step LSTM and Transformer~\citep{hsu2022spatiotemporal}, CMC-COV19D based on contrastive representation learning (CRL)~\citep{hou2022fdvtss}, and 3D ResNets~\citep{Turnbull2022}.

Still in medical imaging field, the convolutional recurrent neural networks method proposed by~\cite{kollias2018deep} detects and predicts Parkinson’s based on medical imaging information.

\section{Methodology}

This section describes the main aspects related to our methodology developed for COVID-19 detection based on a 3D convolutional neural network.

\subsection{Data Processing}

Due to the size of the dataset and our hardware limitations, the images had to be processed before training. Aiming this, we resized every CT scan 3D image, using a spline interpolation to reduce all images to 224$\times$224 pixels, and to make all sample folders have 64 slices.

\subsection{Data Augmentation}

Data augmentation is a process that inflates the dataset, by creating more and more diverse data for the network to learn, when we have a very limited amount of data at hand. In this way, we can use more data to avoid overfitting. Due to the limited amount of samples, data augmentation techniques were used.

The operations of data augmentation used were: additive Gaussian noise, with a mean of 0, and a standard deviation randomly defined from a uniform distribution, ranging from 0 to 20; Gaussian blur, with a mean of 0, and a standard deviation randomly defined from a uniform distribution, ranging from 0 to 2; rotation, with an angle of rotation randomly defined from a uniform distribution, ranging from -30 to 30; flip (vertical and horizontal); cutout, which fills the image with 0 to 4 gray rectangular areas, with height and width that have 20\% of the dimensions of the image; and gamma contrast, with gamma values randomly defined from a uniform distribution, ranging from 0.5 and 2. The operations additive Gaussian noise, Gaussian blur, flip (vertical and horizontal) and gamma contrast are applied randomly, with a rate of 0.5. Moreover, when selected, the operations are applied in a random order.

\subsection{MIA-3DCNN}

\begin{figure}[!htb]
\centering
\subfloat[COVID-19 detection architecture.]{\includegraphics[width=0.99\textwidth]{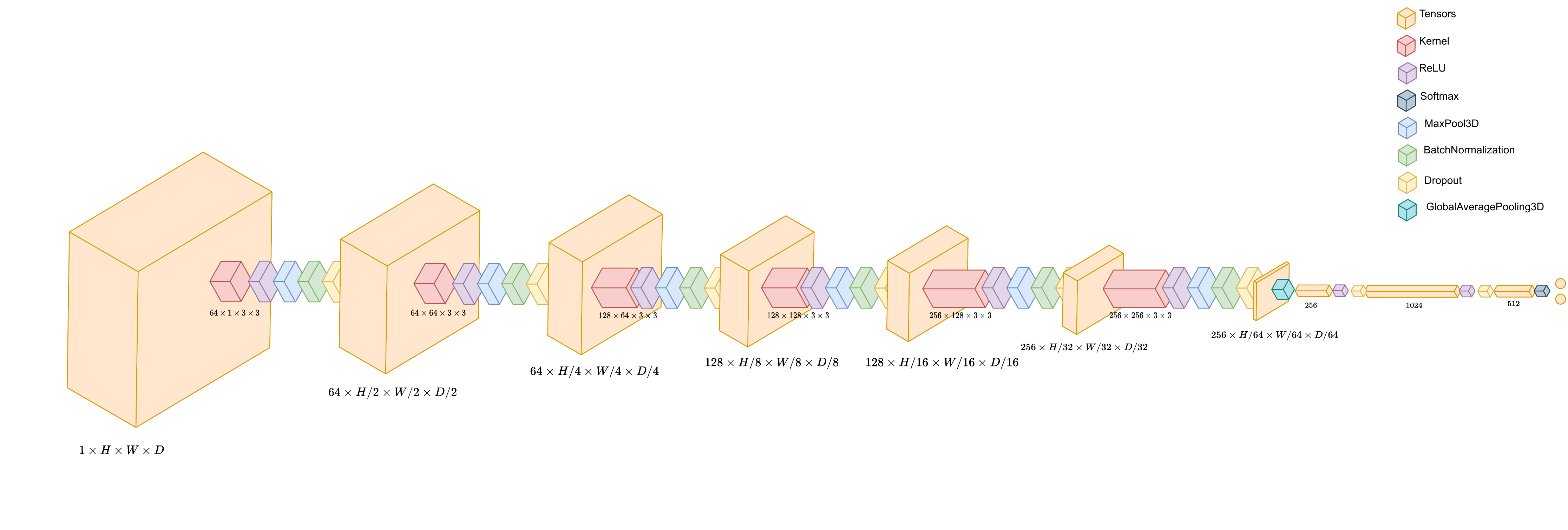} \label{fig:detection-arch}} \\
\subfloat[COVID-19 severity classification architecture.]{\includegraphics[width=0.99\textwidth]{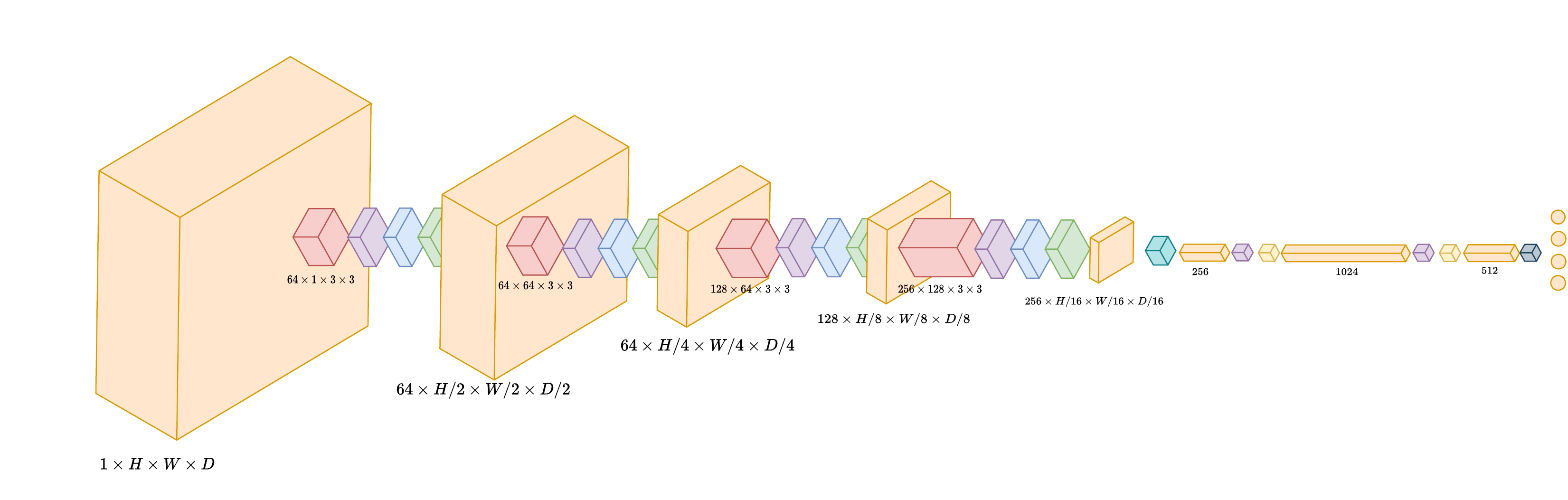} \label{fig:severity-arch}}
\caption{Diagrams for the MIA-3DCNN COVID-19 detection~\protect\subref{fig:detection-arch} and severity~\protect\subref{fig:severity-arch} network architectures used in the challenge, where we used an input CT scan with $H=224$, $W=224$, and $D=64$.}
\label{fig:architectures}
\end{figure}

The method adopted to detect COVID-19 in the images is the MIA-3DCNN network, a 3D convolutional neural network.\footnote{The implementation can be found in
the GitHub repository: \url{https://github.com/igorknz/mia-3dcnn}}. Our proposed architecture has two main stages: one composed of 3D convolutional blocks, and one composed of fully connected layers (Figure~\ref{fig:detection-arch}).

The 3D convolutional stage has blocks composed of one 3D convolutional layer, a 3D max pooling layer, a batch normalization layer and a dropout layer. The 3D convolutional layers have a kernel size of 3. In addition, they have L2 regularizers for the weights and biases, that were added to increase the generalization capability of the network.

The regularizing factor for the biases were kept at 0.01, while the ones for the weights change according to Table \ref{table:cnn-layers}. In addition, the layers use padding, so the output maintains the same dimension of the input. This is specially important, because, the images had to be considerably reduced, to lower the processing required during training.

The layers also use rectified linear unit (ReLU) as the activation function, and He Normal as the weight initializer, to reduce the time of convergence. Then, there is a 3D max pooling layer, with pooling size of 2; a batch normalization layer, and a dropout layer, with rate of 0.5. These are the components of the blocks used for feature extraction. Six of them are stacked at the beginning of the network. The changes between them are the number of filters and the regularization factor for the L2 regularizers, for both weight. These different setups are listed in Table~\ref{table:cnn-layers}, with block number 1 being the one at the bottom of the network and block number 6 being the one close to the top.

\begin{table}[!htb]
\centering
\caption{Parameters that differ along the convolutional blocks of the network.}
\begin{tabular}{ccc}
\toprule
\textbf{Convolutional block} & \textbf{Number of filters} & \textbf{L2 weight factor} \\
\midrule
1 & 64  & 0.01 \\ 
2 & 64  & 0.01 \\ 
3 & 128 & 0.05 \\ 
4 & 128 & 0.05 \\ 
5 & 256 & 0.05 \\ 
6 & 256 & 0.05 \\ 
\bottomrule
\end{tabular}
\label{table:cnn-layers}
\end{table}

After this, there is a 3D global average pooling layer to downsample the output of the initial stage. Then, we have a second stage, composed of blocks that have a fully connected layer with varying number of neurons, according to Table~\ref{table:fc-layers} and a dropout layer, with a rate of 0.5. At last, there is a 2-neuron layer with softmax as the activation function, for the classification, used as the output.

\begin{table}[!htb]
\centering
\caption{Parameters that differ along the convolutional blocks of the network.}
\begin{tabular}{ccc}
\toprule
\textbf{Fully connected block} & \textbf{Number of neurons} \\
\midrule
1 & 1024 \\ 
2 & 512 \\ 
\bottomrule
\end{tabular}
\label{table:fc-layers}
\end{table}

The final architecture was achieved experimentally, after successive experiments. Initially, we decided to use a convolutional neural network due to their great results in finding image patterns in classification tasks. For specific challenge, we decided to use a 3D network due to the structure of the samples. We started by creating a network of stacked 3D convolutional layers for feature extraction, followed by dense layers for classification, and adapted the network after every experiment, by interpreting the metrics obtained.

For the 3rd COVID-19 Competition, we investigated and submitted the MIA-3DCNN network trained both with (version A) and without (version B) data augmentation operations. These respective results can be found in the following section.

During the COVID-19 severity task, another version of MIA-3DCNN (version C) was adopted, using the data augmentation, and is shown in Figure~\ref{fig:severity-arch}. 

Similarly to the previous one, this architecture has a convolutional stage, composed of four blocks of one 3D convolutional layer, a 3D max pooling layer. The specification of each convolutional block is given in Table~\ref{table:cnn-layers-sev}.

\begin{table}[!htb]
\centering
\caption{Parameters that differ along the convolutional blocks of the network.}
\begin{tabular}{ccc}
\toprule
\textbf{Convolutional block} & \textbf{Number of filters} & \textbf{L2 weight factor} \\
\midrule
1 & 64  & 0.05 \\ 
2 & 64  & 0.05 \\ 
3 & 128 & 0.10 \\ 
4 & 256 & 0.10 \\ 
\bottomrule
\end{tabular}
\label{table:cnn-layers-sev}
\end{table}

Subsequently, MIA-3DCNN has the same classification block as versions A and B, consisting of a 3D global average pooling layer and two fully connected layer blocks (Table~\ref{table:fc-layers}), with a dropout (rate of 0.5). Finally, we have a 4-neuron layer output with a softmax activation function for the classification in mild, moderate, severe and critical classes.

\subsection{Other Parameters}

During the training of the models, we used the macro F1 score in the validation set to verify if the model is improving. We used a learning rate scheduler to reduce the learning rate by a factor of 0.5 if the macro F1 score did not improve in the validation set for 20 epochs, with an initial value of $10^{-4}$.

Due to the imbalance in the number of samples in the COVID and non-COVID classes, we used class weights to prevent the network from learning potential biases during the training process. 

Moreover, during COVID-19 detection, we used the categorical crossentropy as the loss function, Rectified Adam as optimizer, batch size of 5, early stopping with a patience of 80 epochs, to reduce the number of iterations required, and the maximum number of epochs was set to 500. On the other hand, for COVID-19 severity classification, SGD optimizer was used with a learning rate of $10^{-4}$, and a momentum of 0.9, and the early stopping was set with a patience of 50 and 1000 as maximum number of epochs.

\section{Results}

This section describes the results obtained with our 3D convolutional neural network.

The final macro F1 scores in the validation set, for both versions submitted for COVID-19 detection task, are listed in Table~\ref{table:covid-class-results}. As we can observe, version~A (with data augmentation) achieved a lower score than version~B (without data augmentation). Even though data augmentation can prevent overfitting, it can also make it more challenging for the network to learn patterns, specially in this case, with randomly applied operations to each of the samples.

\begin{table}[!htb]
\centering
\caption{Performance evaluation of validation set of the proposed method for COVID-19 detection task.}
\begin{tabular}{cc}
\toprule
\textbf{Model} & \textbf{macro F1 Score} \\
\midrule
MIA-3DCNN (Version A) &  0.8681 \\ 
MIA-3DCNN (Version B) &  0.8876 \\ 
\bottomrule
\end{tabular}
\label{table:covid-class-results}
\end{table}

It is expected that the data in the validation set was taken from the same distribution of the data in the training set. The images generated from the data augmentation operations lead to images that are from a distribution that is slightly different from the validation one, which can explain the results obtained.

Regarding the severity classification task, although the severity classification is a challenging task, containing four classes that requires a careful process with specialists to classify them, our model could achieve a considerable result of almost 0.73, as shown in Table~\ref{table:covid-sev-results}.

\begin{table}[!htb]
\centering
\caption{Performance evaluation of validation set of the proposed method for COVID-19 severity classification task.}
\begin{tabular}{cc}
\toprule
\textbf{Model} & \textbf{macro F1 Score} \\
\midrule
MIA-3DCNN (Version C) &  0.7277 \\ 
\bottomrule
\end{tabular}
\label{table:covid-sev-results}
\end{table}

For both challenges, the results surpass the baseline results presented by~\citeauthor{white_paper}, 0.74 and 0.38 macro F1 score for COVID-19 Detection and
COVID-19 Severity Detection validation sets, respectively.

\section{Conclusions}

This paper describes our 3D convolutional neural network for detecting COVID-19 on CT images from the MIA-COV19D database. Although the dataset presents various challenging scenarios, the results obtained are promising and demonstrate the effectiveness of the proposed architecture, surpassing the baseline results in around 15\% for the COVID-19 detection task, and in 35\% for COVID-19 Severity Detection validation set. 

\section*{Acknowledgments}

The authors would like to thank Semantix Brasil for sponsoring the development of this work, and CENAPAD-SP for providing high performance processing environments.

\bibliographystyle{dcu}
\bibliography{references}

\end{document}